\documentclass[useAMS,usenatbib]{mnras}

\usepackage{aas_macros}
\usepackage{apjfonts}
\usepackage{amsfonts}
\usepackage{amsmath}
\usepackage{amssymb}
\usepackage{graphicx}
\usepackage{comment}
\usepackage{natbib}%
\begin{document}

\title{Diffusive photospheres in gamma-ray bursts}

\author[Vereshchagin and Siutsou]{G.~V. Vereshchagin$^{1,2,3,4}$ and I.~A. Siutsou $^{3,5}$ \\
$^{1}${ICRANet, 65122, p.le della Repubblica, 10, Pescara, Italy} \\
$^{2}${ICRA and Dipartimento di Fisica, Sapienza Universit\`a di Roma, P.le Aldo Moro 5, 00185 Rome, Italy} \\
$^{3}${ICRANet-Minsk, National Academy of Sciences of Belarus, Nezavisimosti ave. 68, 220072 Minsk, Belarus} \\
$^{4}${INAF, Viale del Parco Mellini 84, 00136 Rome, Italy} \\
$^{5}${Institute of Physics of Belarus NAS, Nezavisimosti Avenue 68, Minsk 220072, Belarus}}

\maketitle

\begin{abstract}
Photospheric emission may originate from relativistic outflows in two
qualitatively different regimes:\ last scattering of photons inside the
outflow at the photospheric radius, or radiative diffusion to the boundary of
the outflow. In this work the measurement of temperature and flux of the
thermal component in the early afterglows of several gamma-ray bursts (GRBs)
along with the total flux in the prompt phase are used to determine initial
radii of the outflow as well as its Lorentz factors. Results indicate that in
some cases the outflow has relatively low Lorentz factors $\Gamma<10$, favouring cocoon
interpretation, while in other cases Lorentz factors are larger $\Gamma>10$,
indicating diffusive photospheric origin of the thermal component, associated
with an ultrarelativistic outflow.


\end{abstract}

\section{Introduction}

Gamma-ray bursts (GRBs) are strong and short flashes of hard radiation
originating at cosmological distances. Since their discovery a number of
dedicated space observatories and ground based telescopes are constantly
monitoring the sky daily reporting new bursts and measuring
distance to their host galaxies. GRBs come in two kinds: short and long, with
their possible progenitors being binary neutron star mergers and collapsing
massive stars reaching the endpoint of their evolution, respectively. Observed
emission in GRBs is well separated in two distinct episodes:\ brief and highly
irregular prompt phase with dominant hard X-ray and $\gamma-$ radiation, and
smoothly decaying long lasting afterglow emission with broadband spectra,
ranging from radio waves up to sub-TeV\ energies. Extremely large energies released in $\gamma$-rays ( $\leq 10^{54}$ erg) as well as a short variability time ($\leq 10$ ms) point to ultrarelativistic outflows giving rise to the observed emission \citep{2018pgrb.book.....Z}.

Prompt emission spectra are non-thermal, their origin is usually associated
with the synchrotron mechanism in relativistic shock waves
\citep{1994ApJ...430L..93R}. Photospheric models with possible dissipation of
kinetic energy of the outflow are attractive alternative to the synchrotron
models since observation of thermal radiation allows determination of basic
hydrodynamic characteristics of the outflow from which these bursts originate
\citep{2014IJMPD..2330003V,2017IJMPD..2630018P}. The photons in these models are trapped and advected with the outflow until it becomes transparent. In many GRBs subdominant
thermal component was detected during their prompt emission, while in several
GRB 090902B observed spectrum is almost thermal \citep{2010ApJ...709L.172R,2017MNRAS.472.1897R}.

Thermal components are also detected in time resolved spectra during the early
afterglow in a number of GRBs
\citep{2011MNRAS.416.2078P,2012MNRAS.427.2950S,2012MNRAS.427.2965S,2013ApJ...771...15F,2018MNRAS.474.2401V,2019Natur.565..324I}.
So far several mechanisms to generate such emission are proposed. They include
a shock breakout from a progenitor star or a stellar wind
\citep{2006Natur.442.1008C} and a hot cocoon formed when the relativistic jet
emerges from the stellar surface \citep{2006ApJ...652..482P,2017ApJ...834...28N}. Some authors
argue that shock breakouts are not energetic enough and do not last long
enough to explain observed thermal emission \citep{2018MNRAS.474.2401V},
leaving cocoons as a favourite model. In addition, there is an alternative
proposal of a cloud or a clump with small mass, accelerated by the GRB outflow
\citep{2017A&A...600A.131R}.

Most papers dealing with the photospheric emission, e.g.
\citep{2000ApJ...530..292M,2008ApJ...682..463P,2011ApJ...732...49P,2011ApJ...737...68B,2013MNRAS.428.2430L,2016MNRAS.456.1049S,2018ApJ...852...24B}, for a review see \cite{2017IJMPD..2630018P}, adopt the hydrodynamic model of a
steady and infinite wind. However, finite duration of GRBs implies finite
width of the wind. Winds of \emph{finite duration} are classified as photon
thin and photon thick \citep{2013ApJ...767..139B,2013ApJ...772...11R,2014IJMPD..2330003V}.
Decoupling of photons from plasma in the latter case occurs simultaneously in
the entire outflow, while in the former case photons are transported to the
boundaries of the outflow by radiative diffusion, just like in nonrelativistic
outflows, e.g. in supernova ejecta. Emission in this case originates not at
the photospheric radius, but at smaller radii. The photon thick case,
corresponding to the steady wind, appears to be justified for typical GRB
parameters. Photon thin regime is not considered in the literature, as it is
assumed that at large radii the outflow is spreading
\citep{1993MNRAS.263..861P,1993ApJ...415..181M} due to strong velocity
gradients initially present in the outflow, see e.g.
\citep{1993MNRAS.263..861P,1993ApJ...415..181M}. Such spreading outflows
indeed correspond to the photon thick case \citep{2014NewA...27...30R}.
However, in absence of these gradients the outflow could be photon thin where
decoupling of photons from expanding plasma occurs in the diffusive regime \citep{2013ApJ...772...11R}.

Radiative diffusion is known to be relevant for expanding ejecta in supernovae
explosions \citep{1996snih.book.....A}, but was overlooked in the literature
on GRBs. The purpose of the present work is to develop further the theory of
photospheric emission \citep{2013ApJ...772...11R}, specifically focusing on
the case when observed properties of such outflows are determined by the
radiative diffusion of photons, and to confront it with the observational data.

The paper is organized as follows. The definition of the radius of photosphere
is recalled in Section 2. Observational properties of diffusive photospheres
are discussed in Section 3. The method allowing determination of initial
radius and bulk Lorentz factor of the outflow is presented in Section 4.
Observational properties of GRB cocoons are discussed in Section 5. Case
studies of GRBs with thermal emission in the early afterglow is performed in
Section 6. Discussion and conclusion follow. Appendix collects basic results
for ultrarelativistic diffusive photosheres derived from the radiative
transfer theory.

\section{Relativistic photosphere}

Consider a relativistic outflow launched at a radius $R_{0}$. The outflow is
characterized by its activity time $\Delta t$, the luminosity $L$ and mass
injection rate $\dot{M}$. The associated thickness of the outflow is
$l=c\Delta t$. The entropy in the region where the energy is released is
parametrized by a dimensionless parameter $\eta=L/\dot{M}c^{2}$. Spherical
symmetry is assumed, but generalization for anisotropic case with $\eta\left(
\theta\right)  $, where $\theta$ is the polar angle is straightforward. When
$\eta\gg1$ the bulk Lorentz factor changes with the radial distance as%
\begin{equation}
\Gamma\simeq\left\{
\begin{array}
[c]{cc}%
\dfrac{r}{R_{0}}, & R_{0}<r<\eta R_{0},\\
\  & \\
\eta\simeq\mathrm{const}, & r>\eta R_{0},
\end{array}
\right.  \label{gammaeq}%
\end{equation}
During both acceleration and coasting phases the continuity equation for the
laboratory number density reduces to
\begin{equation}
n=\left\{
\begin{array}
[c]{cc}%
n_{0}\left(  \dfrac{R_{0}}{r}\right)  ^{2}, & R(t)<r<R(t)+l,\\
\  & \\
0, & \mathrm{otherwise,}%
\end{array}
\right.  \label{nxi}%
\end{equation}
where $R(t)$ is the radial position of the inner boundary of the outflow.

The optical depth for a spherically symmetric outflow is
\citep{1991ApJ...369..175A,2013ApJ...772...11R}%
\begin{equation}
\tau=\int_{R}^{R+\Delta R}\!\!\!\sigma n\left(  1-\beta\cos\theta\right)
\frac{dr}{\cos\theta}, \label{tau}%
\end{equation}
where $R+\Delta R$ is the radial coordinate at which the photon leaves the
outflow, and $\theta$ is the angle between the velocity vector of the outflow
and the direction of propagation of the photon, $n$ is the laboratory number
density of electrons and positrons, which may be present due to pair
production. The dominant interaction of photons in our case is Compton
scattering in the non-relativistic regime, so $\sigma$\ is the Thomson cross section.

Electron-positron-photon plasma with baryon loading reaches thermal
equilibrium before its expansion starts
\citep{2009PhRvD..79d3008A,2008AIPC..966..191A}. With decreasing entropy
$\eta$ opacity due to electrons associated with baryons increases and
eventually dominates over pair opacity. For the laboratory density profile
(\ref{nxi}) one has in the radial direction%
\begin{equation}
\tau=\left\{
\begin{array}
[c]{cc}%
\dfrac{1}{6}\tau_{0}\left(  \dfrac{R_{0}}{R}\right)  ^{3}, & R_{0}\ll R\ll\eta
R_{0},\\
& \\
\dfrac{1}{2\eta^{2}}\tau_{0}\left(  \dfrac{R_{0}}{R}\right)  , & \eta R_{0}\ll
R\ll\eta^{2}R_{0},\\
& \\
\tau_{0}\left(  \dfrac{R_{0}}{R}\right)  ^{2}, & R\gg\eta^{2}R_{0},
\end{array}
\right.  \label{tau2}%
\end{equation}
where%
\begin{equation}
\tau_{0}=\frac{\sigma L}{4\pi m_{p}c^{3}R_{0}\eta}=n_0 \sigma R_0. \label{tau0}%
\end{equation}
The first two lines correspond to a \emph{photon thick outflow} and the
third line corresponds to a \emph{photon thin outflow} \citep{2013ApJ...772...11R}.

The photospheric radius $R_{ph}$ is defined by equating (\ref{tau2}) to unity%
\begin{equation}
R_{ph}=\left\{
\begin{array}
[c]{cc}%
R_{0}\left(  \dfrac{\tau_{0}}{6}\right)  ^{1/3}, & \tau_{0}\ll\eta^{3},\\
& \\
R_{0}\dfrac{\tau_{0}}{2\eta^{2}}, & \eta^{3}\ll\tau_{0}\ll4\eta^{4}\frac
{l}{R_{0}},\\
& \\
(\tau_{0}R_{0}l)^{1/2}, & \tau_{0}\gg4\eta^{4}\frac{l}{R_{0}}.
\end{array}
\right.  \label{Rph}%
\end{equation}
In eqs. (\ref{tau2}) and (\ref{Rph}) the regions of validity of different
approximations are expressed either for radius or for parameters of the
outflow. The crucial parameter which determines whether the outflow is photon
thick or thin is the ratio%
\begin{equation}
\chi=\frac{\tau_{0}}{4\eta^{4}}\frac{l}{R_{0}}. \label{chidef}%
\end{equation}
The outflow is photon thin for $\chi\gg1$ and it is photon thick otherwise.

\section{Relativistic diffusive photosphere}

The definition of the photosphere implies that at this position in space the outflow as a whole becomes transparent to radiation. However, emission emerges from the outflow when it is optically thick as well. Such emission is due to radiative diffusion, which transfers the energy from deeper parts of the outflow towards its surface. Naively one can think that such an effect is negligible in ultrarelativistic outflows. However, this is not the case \citep{2014IJMPD..2330003V}. Indeed, the comoving time, which photon takes to cross the outflow with comoving thickness $l_c=\Gamma l$ is $t_c=l_c^2/D_c$, where $D_c=c/3\sigma n_c$ is the diffusion coefficient, $n_c=n/\Gamma$ is the comoving density of the outflow. The radial coordinate of the outflow at this time is $R\simeq \Gamma c t_c$.

This diffusion radius is found in \citep{2013ApJ...772...11R}, and it is given by
\begin{equation}
R_{D}=\left(  \tau_{0}\eta^{2}R_{0}l^{2}\right)  ^{1/3}, \label{diffradius}%
\end{equation}
where eq. (\ref{tau0}) has been used. It turns out to be always smaller than the photospheric radius of photon thin
outflow, $R_{D}\ll R_{ph}$, so the radiation escapes such an outflow before it
becomes transparent, just like in the supernova ejecta. In this sense the
characteristic radius of the photospheric emission is not the photospheric
radius found from (\ref{tau2}), but the radius of diffusion (\ref{diffradius}%
). The probability distribution of last scattering of photons in diffusive photospheres is qualitatively different from the usual photospheric emission \citep{2013ApJ...767..139B}. Besides, the comoving temperature of escaping radiation is different from
the temperature at the photospheric radius. Applicability of the photon thin
asymptotics, last line in eq. (\ref{Rph}), can be written using eq.
(\ref{diffradius}) as%
\begin{equation}
l\ll\frac{R_{D}}{\sqrt{2}\eta^{2}}. \label{Diff_app}%
\end{equation}
For larger thickness $l$ photon thin asymptotics disappears, so in the
limit $l\rightarrow\infty$ the stationary wind with photon thick asymptotics
is recovered.

Adiabatic expansion implies \citep{2012arXiv1205.3512R} that the observed
temperature of the outflow does not change while it is accelerating, and it
decreases as $T_{obs}\propto R^{-2/3}$\ at the coasting phase. Taking into
account finite size of emitter and cosmological redshift one has
\citep{2007ApJ...664L...1P}%
\begin{equation}
T_{obs}=\frac{\xi}{1+z}T_{0}\left(  \frac{\eta R_{0}}{R_{D}}\right)  ^{2/3},
\label{Tobs}%
\end{equation}
where $\xi$\ is a numerical factor of order unity, $z$ is cosmological redshift. In estimates below
$\xi=1.48$ is assumed following \citep{2007ApJ...664L...1P}, which is found
from the Monte Carlo simulations in the infinite wind approximation, though
the value of $\xi$ in the acceleration phase and in the photon thin case could
be slightly different. The temperature at the base of the outflow is%
\begin{equation}
T_{0}\simeq\left(  \dfrac{L}{4\pi caR_{0}^{2}}\right)  ^{1/4}, \label{T0}%
\end{equation}
where $a=4\sigma_{SB}/c$\ is the radiation constant, $\sigma_{SB}$\ is the
Stefan-Boltzmann constant. Finally, the duration of photospheric emission for
a distant observer is%
\begin{equation}
t_{a}^{D}=(1+z)\frac{R_{D}}{2\eta^{2}c}. \label{taD}%
\end{equation}
In the photon thick case the duration of thermal emission is determined by the
width of the outflow $l$, which is unconstrained. In the photon thin case this
duration is given by eq. (\ref{taD}) and it is a function of the diffusion
radius and the Lorentz factor.

The luminosity of photospheric component scales with radius as%
\begin{equation}
L_{ph}=L_{0}\left(  \frac{\eta R_{0}}{R}\right)  ^{8/3}, \label{luminosity}%
\end{equation}
and the applicability condition of the photon thin case in eq. (\ref{Rph}) and
together with the definition of diffusion radius in eq. (\ref{diffradius}) imply for the
luminosity of diffusive photosphere%
\begin{equation}
L_{thin}<L_{0}\left(  \frac{R_{0}}{\eta l}\right)  ^{8/3}\ll L_{0}.
\label{Lthin}%
\end{equation}
This means that thermal emission is much weaker than the emission of the
prompt radiation, if $\gamma$-rays are produced with high efficiency there. For
$l\sim R_{0}$ this condition strongly favours small values of $\eta$\ and,
consequently, small Lorentz factors of the outflow.

\section{Determination of initial radius and bulk Lorentz factor of the
outflow}

Assuming that the observed thermal component in early afterglows of some GRBs
is of photospheric origin, one can estimate initial radius of the outflow
directly from observations \citep{2007ApJ...664L...1P}. Indeed, in the
ultrarelativistic regime one has%
\begin{equation}
\mathcal{R}\equiv\left(  \frac{F_{obs}^{BB}}{\sigma_{SB}T_{obs}^{4}}\right)
^{1/2}=\zeta\frac{\left(  1+z\right)  ^{2}}{d_{L}}\frac{R}{\Gamma},
\label{Rcal}%
\end{equation}
where $R$\ is the emission radius, $d_{L}$\ is
the luminosity distance, $\zeta$\ is a numerical factor of
order unity. Following \citep{2007ApJ...664L...1P} $\zeta=1.06$ is assumed for
the estimates below. In \citet{2007ApJ...664L...1P}\ the emission radius $R$ was
associated with the photosphere of the \emph{photon thick} outflows. However,
this relation is valid for any ultrarelativistic emitter. Therefore, from eqs.
(\ref{Tobs}), (\ref{Rcal}) and (\ref{T0}) the initial radius is%
\begin{equation}
R_{0}=\frac{4^{3/2}d_{L}}{\xi^{6}\zeta^{4}\left(  1+z\right)  ^{2}}%
\mathcal{R}\left(  \frac{F_{obs}^{BB}}{YF_{obs}}\right)  ^{3/2}. \label{R0det}%
\end{equation}
In the derivation of this results only two assumptions are made. First, the
outflow should be coasting at ultrarelativistic speed, $\Gamma=\eta\gg1$.
Secondly, the relation $L=4\pi d_{L}^{2}YF_{obs}$\ is used, where $Y$ is the fraction of the total luminosity $L$ and
the energy emitted in X and $\gamma$-rays \emph{in the prompt phase}.

In addition to the initial radius $R_{0}$\ an equation for the Lorentz factor
can be obtained. Since the emitter radius for the \emph{photon thin} outflows
is the diffusion radius $R=R_{D}$, from eqs. (\ref{diffradius}) and
(\ref{Rcal}) one obtains, see also \citep{2014ApJ...792...42B}%
\begin{equation}
\frac{\eta}{l}=\frac{\zeta^{3/2}\left(  1+z\right)  ^{3}}{d_{L}^{1/2}}\left(
\frac{\sigma YF_{obs}}{m_{p}c^{3}\mathcal{R}^{3}}\right)  ^{1/2}.
\label{etaconstr}%
\end{equation}
Therefore, the Lorentz factor can be determined if $l$ is known. In particular case
$l=R_{0}$ from (\ref{etaconstr}) it follows the minimum Lorentz factor for
which the photon thin case applies
\begin{equation}
\eta_{thin}=\left(  1+z\right)  \left(  d_{L}\frac{YF_{obs}\sigma}{m_{p}%
c^{3}\mathcal{R}}\right)  ^{1/2}\frac{4^{3/2}}{\zeta^{5/2}\xi^{6}}\left(
\frac{F_{obs}^{BB}}{YF_{obs}}\right)  ^{3/2}. \label{etadet}%
\end{equation}
The condition (\ref{Diff_app}) determines the applicability limit of the
photon thin case, so for $\sqrt{2}\eta^{2}l=R_{D}$ the photon thick case is
recovered \citep{2007ApJ...664L...1P}%
\begin{equation}
\eta_{thick}=\left[  \zeta\left(  1+z\right)  ^{2}d_{L}\frac{\sigma YF_{obs}%
}{m_{p}c^{3}\mathcal{R}}\right]  ^{1/4}. \label{etaPeer}%
\end{equation}
Equations (\ref{etaconstr}) and (\ref{etadet}) allow the determination of the
bulk Lorentz factor of the \emph{photon thin} outflow, provided the
measurement of the total flux $F_{obs}$ and the parameter $\mathcal{R}$.
Comparing eqs. (\ref{etadet}) and (\ref{etaPeer}) leads to the conclusion that
the Lorentz factor inferred from the photon thin asymptotics is typically
smaller than the one of the photon thick case; besides, it contains the
inverse of the $Y$\ parameter, unlike a factor $Y^{1/4}$ in the latter case.

It is important to stress that from the theoretical point of view, given the
total luminosity and the initial radius of the outflow one cannot distinguish
between photon thick and photon thin cases as both sets of parameters are
possible with different $\eta$ and photospheric radius. These parameters are
related by eq. (\ref{Rcal}), therefore independent observational information
is required in order to differentiate between the two cases.

\section{GRB cocoons}

Consider typical parameters relevant for GRB jets which is penetrating the
progenitor \citep{2017ApJ...834...28N}. In what follows introduce the notation
$A_{x}=A/10^{x}$, so the luminosity $L_{52}$ stands for $10^{52}$ erg/s, which is
the isotropic luminosity. While the entropy of the jet can take large values,
the mixing between the progenitor and the jet lowers the entropy of the
cocoon, so $\eta=10$ is chosen, as a reference value. It is also likely that
the entropy is a decreasing function of the angular distance from the jet.
Assume that initial radius of the wind $R_{0}$ is given by the radius of the
core of the progenitor WR star $R_{0}\sim10^{9}$ cm, and the thickness of the
wind corresponds to the size of the WR star $l_{12}=10^{12}$ cm
\citep{2007ARA&A..45..177C}. The crucial parameter which determines whether
the outflow is photon thick or thin is the ratio%
\begin{equation}
\chi\simeq29L_{52}l_{12}^{-1}\eta_{1}^{-5}. \label{chipar}%
\end{equation}
For $\chi\gg1$ the outflow is photon thin, which is the case for our fiducial
parameters. Considering the extreme dependence on $\eta$, for smaller $\eta
$\ the condition is clearly satisfied. Hence the cocoon is in the photon thin
regime and therefore the radiation from the cocoons is governed by radiative
diffusion. The diffusion radius is%
\begin{equation}
R_{D}\simeq4.9\times10^{14}L_{52}^{1/3}\eta_{1}^{1/3}l_{12}^{2/3}\text{ cm,}
\label{rdiffpar}%
\end{equation}
and the arrival time corresponding to this radius is%
\begin{equation}
t_{a}^{D}=(1+z)\frac{R_{D}}{2\eta^{2}c}\simeq81.7(1+z)L_{52}^{1/3}\eta
_{1}^{-5/3}l_{12}^{2/3}\text{ s}, \label{tardiff}%
\end{equation}
which is the typical duration of thermal emission observed in early afterglows
of GRBs.

The observed temperature at the diffusion radius is%
\begin{equation}
T_{obs}=0.12L_{52}^{1/36}R_{9}^{1/6}\eta_{1}^{4/9}l_{12}^{-4/9}\text{keV,}
\label{Tdobs}%
\end{equation}
which is also a typical temperature of thermal emission in early afterglows of
GRBs \citep{2018MNRAS.474.2401V}.

Such inferred values of temperature and duration call for closer attention to
the radiation properties of photon thin outflows.

\section{Case studies}

All GRBs reported in \citet{2018MNRAS.474.2401V} with measured redshifts and
thermal component detected in their early afterglows were considered, namely
GRBs 060218, 090618, 101219B, 111123A, 111225A, 121211A, 131030A, 150727A,
151027A. Observed temperature, thermal flux and total flux were averaged for
the entire duration of the thermal emission. The initial radius was found from
eq. (\ref{R0det}). Two values of the Lorentz factor in photon thin, eq.
(\ref{etadet}), and photon thick, eq. (\ref{etaPeer}), cases were determined.
Then the minimum value of the $Y$\ parameter is found which allows the
duration of the photospheric emission (\ref{taD}) to be not less than the
observed duration of the thermal component. The values are reported in Tab.
\ref{Tab1}. Only six cases allow both photon thick and photon thin
interpretations for the photosphere; for other cases photon thin case does not
apply because eq. (\ref{etaPeer}) gives smaller Lorentz factor than eq.
(\ref{etadet}).

\textbf{GRB 060218.} This is a well studied nearby GRB \citep{2006Natur.442.1008C} with record breaking duration
of the thermal signal interpreted as the break out of a shock driven by a
mildly relativistic shell into the dense wind surrounding the progenitor, see
however \citep{2007MNRAS.382L..77G,2019MNRAS.484.5484E}. The thermal emission
in this burst with observed temperature $T_{BB}=0.15$ keV may be also explained as a photosphere of a cocoon launched from
initial radius $3.18\times10^{11}$ cm with a mildly relativistic Lorentz factor
$1.2Y^{-1}<\Gamma<1.6Y^{1/4}$ emitting in the photon thin regime. This estimate of the Lorentz factor is in agreement with radio observation at 2 days, requiring $\Gamma\sim 2$ \citep{2006Natur.442.1014S}. Given that the condition $\Gamma\gg1$ is not satisfied, results of the theory of diffusive ultrarelativistic photospheres can be applied to this case with great care. In particular, the estimated duration of the thermal signal is only $13Y$ s.

\textbf{GRB 090618.} This burst may represent a canonical case of photon thin
outflow launched from the initial radius $10^{9}$ cm with the Lorentz factor
$3Y^{-1}<\Gamma<40Y^{1/4}$. Assuming instantaneous energy injection with
$l=R_{0}$ one finds $Y=5.7$. The duration of the thermal emission with observed temperature about $1$ keV is about 6$Y$ s. Note that thermal emission has also been claimed in the prompt phase, with a higher temperature ranging from $54$ to $12$ keV
\citep{2012A&A...543A..10I}. Such thermal emission in the prompt phase may be
interpreted as a photosphere of the photon thick outflow. Indeed, if the
initially high entropy $\eta$\ decreases with time the outflow should experience a
transition from photon thick to photon thin case.

\textbf{GRB 111225A.} This case is similar to GRB 060218, but with smaller
initial radius of $8.3\times10^{9}$ cm and Lorentz factor in the range
$1<\Gamma<6.0Y^{1/4}$. The duration of the thermal emission with observed temperature of $0.18$ keV is about 126$Y$ s, with $Y=2.0$ for $l=R_0$. The lower bound on the Lorentz factor in unconstrained. It may correspond to a cocoon emitting in the
diffusive photon thin regime.

\textbf{GRB 131030A.} This case is a typical long burst, similar to GRB
090618, it allows Lorentz factors in the photon thin case in the following
range:\ $4.3Y^{-1}<\Gamma<66Y^{1/4}$. The duration of the thermal emission with observed temperature of $1.12$ keV is about 2.0$Y$ s. The initial radius is $3.74\times 10^{8}$ cm. For instantaneous energy injection $Y=19$.

\textbf{GRB 150727A.} This case is similar to GRB 111225A with initial
radius $1.1\times10^{9}$ cm and Lorentz factors in the range $1<\Gamma<11Y^{1/4}$, allowing for a photon thin interpretation. The duration of the thermal emission with observed temperature of $0.47$ keV is about 191$Y$ s, with $Y=2.1$.

\textbf{GRB 151027A.} This case is similar to GRBs 090618 and 131030, however with quite large initial
radius $1.5\times10^{10}$ cm and Lorentz factors in the narrow range $23<\Gamma<26Y^{1/4}$. The duration of the thermal emission with observed temperature of $0.96$ keV is about 0.47$Y$ s, with $Y=64$.

\section{Discussion}

The difference between the present work and the approach followed by \citet{2006ApJ...652..482P} should be emphasized. The main assumption of that work is the presence of unknown dissipation mechanism, which transforms part of the kinetic energy of the outflow into radiation, postulated in \citep{2005ApJ...628..847R}. Such dissipation can boost the luminosity of thermal emission and it might be required to explain subdominant thermal component during the prompt emission or even the prompt emission itself, see \citep{2020MNRAS.491.4656B}. Concerning observations of this component in the early afterglow, dissipation is not required, as it is much weaker than the prompt radiation.

Similarly, there is a difference between the present work and the work by \citet{2017ApJ...834...28N}. There two regimes of expansion are considered: Newtonian ($v<c$) and ultrarelativistic $\eta\gg 1$. For the latter, which is of interest here, the emission was assumed to originate at the photospheric radius, given by the last line in eq. (\ref{Rph}). As discussed in Sec. 3 above, photons in this case diffuse out much earlier, so that no photons are left in the outflow when it arrives to the photospheric radius. For this reason estimations of the luminosity and observed temperature in that paper cannot be used.

Qualitative difference in dependence of observed flux and temperature on time
for photon thin outflows determine their observed properties. In particular,
since the flux up to diffusion time (\ref{tardiff}) is almost constant, see
eq. (\ref{obflux}) in Appendix, and its luminosity is much weaker than the
prompt radiation luminosity, see eq. (\ref{Lthin}), the thermal component
become visible after the steep decrease of observed luminosity following the
end of the prompt phase. This is indeed where such thermal component is
identified in many GRBs. Its disappearance is naturally explained by diffusion
of the radiation kept in the outflow. Hence it implies that no more photons
are generated neither in the outflow nor in the central engine.

The results of the present work indicate that in several GRBs, namely GRB 060218, 111225A and
150727A, the thermal component observed in the early afterglow may originate
from mildly relativistic cocoons emerging from the progenitors together with the jet, due to
relatively small values of inferred Lorentz factors $\Gamma<10$. At the
same time, such emission observed in GRBs 090618, 131030A and 151027A correspond to
large Lorentz factors $\Gamma>10$, indicating a jet origin of the
photospheric emission. Besides, these results suggest that the progenitors of
some long GRBs, in particular GRB 090618 and 131030A could be rather compact
objects, with radius $l\sim10^{9}$ cm.

It is important to stress that the relatively low temperature of the thermal component observed in the early afterglow with $T\sim0.1-10$ keV, in
contrast with typical temperatures detected during the prompt emission with
$T\sim10-100$ keV does not indicate small Lorentz factor of the outflow.
Conversely, it may point to photospheric origin of the thermal emission in the
photon thin regime. Instead, large Lorentz factors $\Gamma\gg1$ assumed in the
model imply small mass of the emitting plasma, which is consistent with the
cocoon interpretation \citep{2017ApJ...834...28N}.

Possible presence of thermal components both in the prompts radiation and in
the early afterglow, as well as the presence of breaks in temperature
dependence on time found in many cases
\citep{2004ApJ...614..827R,2005ApJ...625L..95R,2009ApJ...702.1211R} may correspond to the
transition from photon thick to photon thin asymptotics in hydrodynamic
evolution of the outflow powering GRBs.


\section{Conclusions}

The theory of diffusive emission from relativistic photospheres is
developed and confronted with observational data on a sample of GRBs with
thermal component in the early afterglows. The measurement of temperature and
flux of the thermal component along with the total flux in the prompt phase
are used to determine initial radii of the outflows as well as their Lorentz
factors.

The results indicate that in several cases (GRBs 060218, 111225A and
150727A) the inferred Lorentz factors are relatively small, $\Gamma<10$, while
in other cases (GRBs 090618, 131030A and 151027A) the inferred Lorentz factors are
larger, $\Gamma>10$. Such differences suggest two possible sources of
the thermal component: mildly relativistic cocoons or highly relativistic jets. This is valid only for those cases, where inferred Lorentz factor is relatively small, below few tens. For other cases identified in \citet{2018MNRAS.474.2401V} inferred Lorentz factors are larger, and photon thin interpretation does not apply.

These results are the first indication that radiative diffusion may play an
important role not only in nonrelativistic outflows, but also in
ultrarelativistic outflows, represented by GRBs.

\vspace{.3in}

{\bf Acknowledgements.} It is a pleasure to thank Dr. Damien B\'egu\'e for careful reading of the manuscript and useful comments.

\bibliographystyle{mnras}

\appendix

\onecolumn

\section{Initial radius and Lorentz factors in
photon thick and photon thin cases}

In Tab. \ref{Tab1} results of calculations of the initial radius $R_0$ and Lorentz factors $\eta$ in
photon thick and photon thin cases. The sample of GRBs with thermal component detected in the early afterglows of GRBs is adopted from \citep{2018MNRAS.474.2401V}. Both flux and observed temperature reported in online material of that paper for time resolved intervals are averaged on the entire duration of observation of thermal component.

\begin{table}
\caption{Results of calculation of Lorentz factors and initial radii for the
set of GRBs from \citet{2018MNRAS.474.2401V}. Reported are: GRB name,
redshift, fluence, duration of the prompt emission $T_{90}$, duration of the
thermal component $\Delta t_{BB}$, estimated duration of the photospheric
emission $\Delta t$, average observed temperature $T_{BB}$, Lorentz factor of
the photon thick case, Lorentz factor of the photon thin case, initial radius
$R_{0}$ and the references: [1] \citep{2006GCN..4822....1S,2006GCN..5376....1F}; [2] \citep{2009GCN..9534....1S,2009GCN..9518....1C}; [3] \citep{2011GCN.12726....1B,2014GCN.16079....1T}; [4] \citep{2013GCN.15456....1B,2013GCN.15407....1X}; [5] \citep{2015GCN.18080....1T,2015GCN.18086....1S}; [6] \citep{2015GCN.18496....1P,2015GCN.18487....1P}.}%
\centering
\begin{tabular}
[c]{llllllllllll}\hline
GRB & $z$ & Fluence, & $T_{90}$, & $\Delta t_{BB}$, &
$\Delta t$, & $T_{BB}$, & $\eta_{thick}$ & $\eta_{thin}$ & $R_{0}$, &
$Y$ & Reference\\
    & & erg/cm$^{2}$ & s & s & s & keV &  &  & $10^{10}$ cm & & \\ \hline
060218 & $0.033$ & $6.8\times10^{-6}$ & $>2000$ & $2624$ & $13Y$ & $0.146$ &
$1.6Y^{1/4}$ & $1.2Y^{-1}$ & $32Y^{-3/2}$ & $206$ & [1] \\
090618 & $0.54$ & $2.7\times10^{-4}$ & $113.2$ & $34$ & $6.0Y$ & $1.05$ &
$40Y^{1/4}$ & $3.0Y^{-1}$ & $0.094Y^{-3/2}$ & $5.7$ & [2] \\
111225A & $0.297$ & $1.3\times10^{-6}$ & $106.8$ & $331$ & $126Y$ & $0.18$ &
$6.0Y^{1/4}$ & $>1$ & $0.83Y^{-3/2}$ & $2.0$ & [3] \\
131030 & $1.293$ & $6.6\times10^{-5}$ & $41$ & $90$ & $2.0Y$ & $1.12$ &
$66Y^{1/4}$ & $4.3Y^{-1}$ & $0.037Y^{-3/2}$ & $19$ & [4] \\
150727A & $0.313$ & $7.9\times10^{-6}$ & $88$ & $518$ & $191Y$ & $0.25$ &
$11Y^{1/4}$ & $>1$ & $0.11Y^{-3/2}$ & $2.1$ & [5] \\
151027A & $0.81$ & $1.94\times10^{-5}$ & $130$ & $55$ & $0.47Y$ & $0.96$ &
$26Y^{1/4}$ & $23Y^{-1}$ & $1.5Y^{-3/2}$ & $64$ & [6] \\ \hline
\end{tabular}
\label{Tab1}%
\end{table}

\section{Emission from photon thin outflows}

\begin{figure}
\centering
\includegraphics[width=.5\columnwidth]{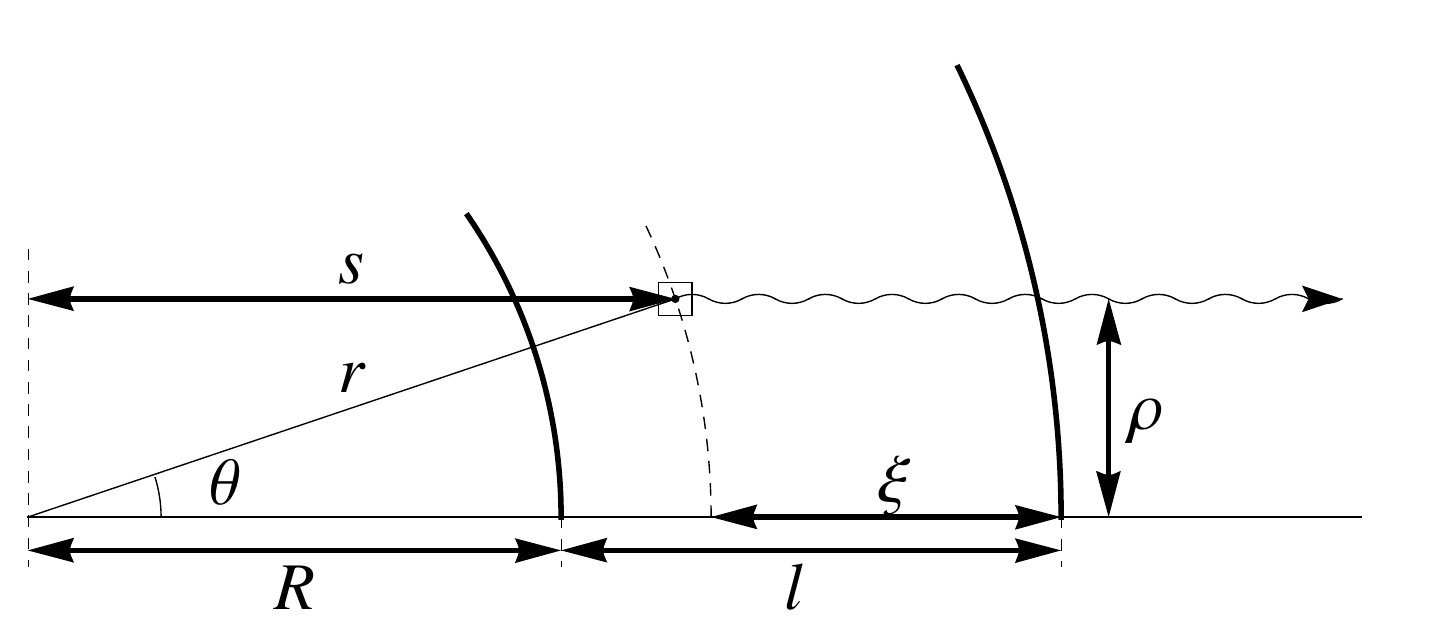}\caption{Geometry of the outflow and
variables used. Observer is located to the right at infinity. }%
\label{xi}%
\end{figure}

Recall the solution of the radiative transfer equation for photon thin
outflows in diffusion approximation \citep{2013ApJ...772...11R}. The radiative
transfer equation for specific intensity $I_{\nu}$ along the ray (see e.g.
\citep{1979rpa..book.....R}, p.~11) is%
\begin{equation}
\frac{dI_{\nu}}{ds}=j_{\nu}-\kappa_{\nu}I_{\nu}, \label{radtransf}%
\end{equation}
where $j_{\nu}$ is monochromatic emission coefficient for frequency $\nu$,
$\kappa_{\nu}$ is absorption coefficient and $s$ is distance, measured along
the ray, see Fig.~\ref{xi}.

Spectral intensity of radiation at infinity on a ray coming to observer at
some arrival time $t_{a}$ is given by formal solution of this equation
\citep{2011ApJ...737...68B}
\begin{equation}
I_{\nu}(\nu,\rho,t_{a})=\int\mathcal{I}_{\nu}(\nu,r,\theta,t)\exp[-\tau
(\nu,r,\theta,t)]\,d\tau, \label{sourcefunc}%
\end{equation}
where $\mathcal{I}_{\nu}(r,\theta,t)=j_{\nu}/\kappa_{\nu}$ is the source
function and the optical depth is%
\begin{equation}
\tau=\int_{s}^{\infty}\kappa_{\nu}ds, \label{taukappa}%
\end{equation}
and variables $(r,\theta,t)$ are connected by $t_{a}=t-(r/c)\cos\theta$ and
$\rho=r\sin\theta$, see Fig. \ref{xi}.

The total observed flux is%
\begin{equation}
F_{\nu}(\nu,t_{a})=2\pi\Delta\Omega\int I_{\nu}(\nu,\rho,t_{a})\,\rho
\,d\rho\,, \label{Fnu}%
\end{equation}
where $\Delta\Omega$ is the solid angle of the observer's detector as seen
from the outflow in the laboratory frame and $2\pi\rho d\rho$ is an element of
area in the plane of the sky.

That emissivity $j_{\nu}$ is assumed to be thermal and isotropic in comoving
frame and $\kappa_{\nu,c}=\mathrm{const}$. The laboratory source function is
then
\begin{equation}
\mathcal{I}_{\nu}(\nu,r,\theta,t)=\frac{2h}{c^{2}}\frac{\nu^{3}}{\exp\left(
\frac{h\nu\Gamma(1-\beta\cos\theta)}{kT_{c}(r,t)}\right)  -1},
\end{equation}
where $h$ is the Planck constant. The source function $\mathcal{I}$ depends on
both $r$ and $t$. The Rosseland radiative diffusion approximation is used (see
e.g. \citep{1979rpa..book.....R}, pp. 39--42). It is useful to introduce the
function $L_{c}(\xi,t)=(t/t_{0})^{8/3}I_{c}(\xi,t)$, which accounts for the
adiabatic cooling of radiation in expanding outflow. Here both $\xi$ and time
$t$ are measured in the laboratory frame, while $I_{c}(\xi,t)$ is measured in
the comoving frame. By applying multipolar decomposition the diffusion
equation was derived from (\ref{radtransf}) in the ultrarelativistic limit
\citep{2013ApJ...772...11R}
\begin{equation}
\frac{\partial L}{\partial ct}-\frac{c^{2}t^{2}\Delta}{3R_{0}}\frac
{\partial^{2}L}{\partial\xi^{2}}=0,\qquad\Delta=\frac{1}{\Gamma^{2}\tau_{0}}.
\end{equation}
Notice that the diffusion coefficient is explicitly time dependent due to the
expansion of the outflow.

This equation should be supplemented with boundary conditions. There are two
types of boundary conditions used frequently: free-streaming, for example in
two-stream approximation (\citep{1979rpa..book.....R}, pp. 42--45), and zero
boundary conditions, that can be used as replacement for free-streaming for
``extrapolated boundary'' \citep{1994JOSAA..11.2727H}. The position of
``extrapolated boundary'' is found as $\xi=-k\frac{c^{2}t^{2}\Delta}{R_{0}}$
($k$ is a constant of order unity, dependent on the approximation used for
free-streaming description), and for the main part of emission it is very close to the
real boundary. In the case of zero boundary conditions $L|_{\xi=0}%
=L|_{\xi=l}=0$ there is a series expansion of solution, that for initial
conditions $L(\xi,t_{0})=1$ gives
\begin{equation}
L_{c}(\xi,t)  =\sum_{n=0}^{\infty}\frac{4}{(2n+1)\pi}\exp\left[
-\frac{\Delta(2n+1)^{2}\pi^{2}c^{3}(t^{3}-t_{0}^{3})}{9R_{0}^{3}}\right]\times\sin\left[  \frac{(2n+1)\pi\xi}{l}\right].
\label{zeroboundsol}%
\end{equation}
This solution in comparison with numerical one with free-streaming boundary
conditions is accurate to a few percent. The corresponding flux is
characterized by an initial burst and then tends to the asymptotic solution,
that corresponds to $t_{0}=0$. with the flux
\begin{equation}
F_{c}(t)=\frac{4t^{2}}{3t_{D}^{2}}\,\,\vartheta_{2}\left[  0,\exp\left(
-\frac{4\pi^{2}}{9}\left(  \frac{t}{t_{D}}\right)  ^{3}\right)  \right]  ,
\label{asympsol}%
\end{equation}
where $\vartheta_{2}$ is the Jacobi elliptic theta function. The raising part
of the corresponding flux of $L_{c}$ through the external boundary of the
outflow scales as $t^{1/2}$, while its decaying part is quasi-exponential one.
The peak of the flux is near the diffusion time
\begin{equation}
t_{D}=\frac{R_{0}}{c}\Delta^{-1/3}, \label{td}%
\end{equation}
and "extrapolated boundary" $\xi=-kR_{0}\Delta^{-1/3}\ll R_{0}$ is very close
to real one as $\Delta\ll1$, that ensures the accuracy of (\ref{zeroboundsol}%
). For practical purpose it is possible to write a simpler expression%
\begin{equation}
F_{c}(t)=\frac{9}{8}\left(  \frac{t}{t_{D}}\right)  ^{1/2}\,\,\exp\left[
-\frac{4}{9}\left(  \frac{t}{t_{D}}\right)  ^{4}\right]  , \label{comflux}%
\end{equation}
which accurately describe both raising and decaying parts of the flux. The
observed flux as function of arrival time is obtained from (\ref{asympsol}) by
integrating over the emitting surface, which gives a factor $(t/t_{0})^{2} $,
and by correcting for adiabatic factor, which gives additional factor
$(t/t_{0})^{-8/3}$, so the final expression is%
\begin{equation}
F_{obs}(t_{a})=\frac{9}{8}\left(  \frac{t}{t_{a}^{D}}\right)  ^{-1/6}%
\,\,\exp\left[  -\frac{4}{9}\left(  \frac{t}{t_{a}^{D}}\right)  ^{4}\right]  .
\label{obflux}%
\end{equation}
This result is in sharp contrast with strongly decreasing flux from
photon thick outflows $F_{obs}(t_{a})\propto t_{a}^{-2}$ \citep{2011ApJ...732...49P,2013ApJ...772...11R}.

The comoving temperature of radiation on the photosphere is determined by the
balance between the energy diffusion from the interior of the outflow and
radiative losses and it is much smaller than the temperature in the interior.
The variation of observed temperature across the emitting surface is small and
hence the observed instantaneous spectrum is very close to the thermal one and
peaks near the observed temperature on the line of sight. Observed temperature
is determined from the observed flux (\ref{obflux}) as $F\propto R^{2}T^{4}$
and the result is%
\begin{equation}
T_{obs}(t_{a})=T_{D}\left(  \frac{t}{t_{D}}\right)  ^{-13/24}\,\,\exp\left[
-\frac{t}{t_{a}^{D}}\right]  , \label{Tobsd}%
\end{equation}
which reflects the fact that for $t>t_{D}$ the observed temperature decreases
exponentially. This result is in contrast with a power law decrease of observed
temperature of photon thick outflows $T_{obs}(t_{a})\propto t_{a}^{-2/3}$ \citep{2011ApJ...732...49P,2013ApJ...772...11R}.

\end{document}